\documentclass[fleqn,10pt,twocolumn]{wlscirep}
\usepackage[utf8]{inputenc}
\usepackage[T1]{fontenc}
\usepackage[switch]{lineno}
\usepackage{graphicx}
\usepackage{dcolumn}
\usepackage{bm}
\usepackage{hyperref}
\usepackage{amsmath}
\usepackage{amssymb}
\usepackage{latexsym}
\usepackage{epsfig}
\usepackage{amsbsy}
\usepackage{array}
\usepackage{amssymb}
\usepackage{setspace}
\usepackage[justification=centerlast]{caption}
\usepackage{xcolor}
\usepackage{pdfpages}
\usepackage{afterpage}

\newcommand\blankpage{%
	\null
	\thispagestyle{empty}%
	\addtocounter{page}{-1}%
	\newpage}

\title{Scale-Free Networks beyond Power-Law Degree Distribution}

\author[1,2]{Xiangyi Meng}
\author[3,*]{Bin Zhou}
\affil[1]{Network Science Institute and Department of Physics, Northeastern University, Boston, Massachusetts 02115, USA}
\affil[2]{Department of Physics and Astronomy, Northwestern University, Evanston, Illinois 60208, USA}
\affil[3]{School of Management Science and Engineering, Nanjing University of Information Science and Technology, Nanjing, 210044, China}

\begin{abstract} 
Complex networks across various fields are often considered to be scale free---a statistical property usually solely characterized by a power-law distribution of the nodes' degree $k$. However, this characterization is incomplete. In real-world networks, the distribution of the degree--degree distance $\eta$, a simple link-based metric of network connectivity similar to $k$, appears to exhibit a stronger power-law distribution than $k$. While offering an alternative characterization of scale-freeness, the discovery of $\eta$ raises a fundamental question: do the power laws of $k$ and $\eta$ represent the same scale-freeness? To address this question, here we investigate the exact asymptotic {relationship} between the distributions of $k$ and $\eta$, proving that every network with a power-law distribution of $k$ also has a power-law distribution of $\eta$, but \emph{not} vice versa. This prompts us to introduce two network models as counterexamples that have a power-law distribution of $\eta$ but not $k$, constructed using the preferential attachment and fitness mechanisms, respectively. Both models show promising accuracy by fitting only one model parameter each when modeling real-world networks. Our findings suggest that $\eta$ is a more suitable indicator of scale-freeness and can provide a deeper understanding of the universality and underlying mechanisms of scale-free networks.

\end{abstract}

\begin{document}


\maketitle

A variety of real-world complex systems, from virtual social platforms to physical neuronal connectomes, can be modelled as complex networks~\cite{netw-introd}. 
Although the precise definition may vary, a complex network is typically a graph or network (as physicists call it) with the nontrivial complexity that is not present in simple, regular graphs like lattices but is often present in real systems, described by some general, appealing laws~\cite{stat-of-netw_ab02}. One of such laws is the \emph{scale-freeness}, denoting the self-similarity of a complex system across various  scales~\cite{complex-crit}. When the term "scale-free" was first {adopted} to networks~\cite{barabasi-albert_ba99}, it was used to denote the unique property of a {power-law} probability distribution of some variable $k$, $f(k)\propto k^{-\alpha}$, such that $f(k)$ is {invariant (free) under the continuous scale transformation $k\to k+\epsilon k $}. This is reminiscent of the idea of renormalization group developed in the statistical field theory~\cite{stat-field-theor-1,stat-field-theor-2}. In contrast to statistical field theory, however, where a ``scale'' explicitly refers to system size, the "scale" $k$ in networks refers to {connectivity}, usually specified as the degree of each node, i.e.,~the number of links connected to the node. A network with more high-degree nodes has higher connectivity. A power-law degree distribution (DD) allows us to make "scale-free" statements on the relative abundance of high-degree nodes, such as "nodes with triple the normal connectivity are half as common as nodes with normal connectivity." The actual value of "normal connectivit" is irrelevant---whether it is a hundred or a million degrees~\cite{scale-free_t05}.

The concept of scale-free networks has since seen significant theoretical advancements in the last two decades~\cite{comment-to-barabasi-albert_ah00,krapivsky2001degree,netw-self-similar_shm05,evans2005scale,crit-phenom-netw_dgm08,szell2010multirelational,lambiotte2019networks,power-law-outage_nsz20}. Due to their simplicity and general applicability, scale-free networks have been adopted as a testing ground for theories such as percolation~\cite{netw-percolation_ceah00} or dynamical system~\cite{netw-resil_gbb16} on networks. An example is percolation on networks that have a power-law DD, which predicts that when the power-law exponent $\alpha$ is less than $3$, a cluster spanning the whole network will almost always exist, regardless of the fraction of nodes that break down. Applying the result to the Internet (for which $\alpha\approx2.5$~\cite{barabasi-albert_ba99}), one can estimate that over $99\%$ of the nodes must be destroyed before the spanning cluster collapses~\cite{netw-percolation_ceah00}, demonstrating the extreme robustness of the Internet under random attacks.

When applying the theoretical results to the empirical realm, however, there is a caveat: an ideal, continuous power-law DD would not be normalizable in the full domain $k\in[0,\infty)$. Thus, the DD cannot be truly power law but requires a small-$k$ ``ultraviolet'' (UV) cutoff, in terms of either a $k_{\min}$ (the minimum degree each node can have) or other nontrivial corrections near a finite positive $k$. The DD  only regains its scale invariance asymptotically in the ``infrared'' (IR) limit as $k\to \infty$. Therefore, a finite-size network can only be approximately scale-free~\cite{finite-size_scmrsbc21}. This raises concerns about whether real-world networks, which are finite by nature, can still be considered as ``scale-free'' networks. 

Real-world networks from diverse fields, such as biology~\cite{goh2002classification,protein-interact_klswpsbkkhcvb19}, social science~\cite{garcia2013social,myers2014information}, and information technology~\cite{goh2001universal,chen2002origin}, have been commonly regarded as scale-free, mostly based on the observation of their power-law DD. This supports the claim that "scale-free networks are universal~\cite{barabasi-albert_ba99}." Yet, this claim remains a topic of debate in recent literature~\cite{power-law-is-rare_bc19, rare-and-everywhere_h19,voitalov2019scale,artico2020rare}. The controversy is rooted in the challenge of accurately determining the range of $k$ values to be used for fitting a power law, which requires prior knowledge of the small-$k$ UV cutoff. Fitting a range of $k$ values that is either too large or too small leads to substantial inaccuracies~\cite{power-law-fit-range_csn09}. One approach to simplify the analysis is to statistically test the full domain of $k$, ignoring possible UV cutoff. However, this oversimplified approach resulted in that more than two thirds of real-world networks do not have a statistically significant power-law DD~\cite{degree-degree-distance_zms20}.

To circumvent the challenge of determining the fitting range, an alternative solution was proposed~\cite{degree-degree-distance_zms20}. Instead of the degree $k$, a link-based metric, the \emph{degree--degree distance}~\cite{degree-degree-distance_zms20}, was adopted. The degree--degree distance $\eta$ is defined for every link $i\leftrightarrow j$ that connects two nodes $i$ and $j$, given by
\begin{equation}
\label{eq_eta}
\eta=\exp \left|\ln k_i-\ln k_j\right|,
\end{equation}
which is determined solely by the network topology. Unlike $k$, which measures the degree of connectivity, $\eta$ measures the {imbalance} of connectivity. Increased presence of high-$\eta$ links (i.e.,~high-degree nodes adjacent to low-degree nodes) corresponds to more imbalanced connectivity in the network, which can lead to higher heterogeneity or lower flow capacity. The characteristics of $\eta$ have since been linked to broader topics~\cite{farzam2020degree,wang2021self,maren20212,log-degree_ec22}, such as network assortativity~\cite{farzam2020degree} and closeness~\cite{log-degree_ec22}.
In addition, many node-based metrics that were originally defined through $k$, such as centrality and clustering coefficient, can be redefined  or generalized using $\eta$, offering new possibilities for designing network metrics.

When fitting to power laws, $\eta$ is found to be less susceptible to the UV cutoff. It turned out that around $60\%$ of real-world networks have a statistically significant power-law degree--degree distance distribution (DDDD) when fitting the full range of $\eta$~\cite{degree-degree-distance_zms20}. The higher statistical significance hints that many real-world networks can, indeed, be considered as scale free, when characterized by the power law of DDDD instead of DD. This finding raises a fundamental question: \emph{do the power laws of DD and DDDD represent the same scale-freeness?} Or, more generally speaking, is there only one scale to determine whether a network is scale-free or not? If the answer is affirmative, then the power law of DDDD, $g(\eta)\sim \eta^{-\beta}$, would just act as a ``proxy'' for the power law of DD, $f(k)\sim k^{-\alpha}$, both representing the same underlying scale-free property of complex networks. 

Here, however, we show that the power law of DDDD is {more than} a "proxy." Our main result [Eq.~\eqref{eq_main}] is that the set of networks with an asymptotic power-law DD is a {proper subset} of those with an asymptotic power-law DDDD. When both DD and DDDD are power laws, we find an exact relationship between their power-law exponents, $\beta=\alpha-1$; meanwhile, we also find that there are networks whose DD is not power law, but DDDD is. Our result indicates that the current understanding of scale-free networks, which is based solely on the power law of DD, is incomplete. 

Recently, there have been an increasing number of scholars~\cite{rare-and-everywhere_h19,degree-degree-distance_zms20,serafino2021true} advocating for the attribution of the scale-free property to a network mechanism, rather than a specific metric. Two extensively discussed scale-free mechanisms in the literature~\cite{netw-mech_lb21} are the \emph{preferential attachment}~\cite{barabasi-albert_ba99} and the \emph{fitness}~\cite{quench-fitness_ccdm02} mechanisms. Interestingly, we will show that by employing either of these mechanisms, network models that do not have a power-law DD but have a power-law DDDD can be built. The resulted counterexamples, representing scale-free networks that go \emph{beyond} having a power-law DD, show promising accuracy when modelling real-world networks.

\section*{Results}

Our main result can be written as
\begin{equation}
\label{eq_main}
\mathcal{D}^2|_\text{power-law} 	
\subsetneq
\mathcal{D}^4|_\text{power-law},
\end{equation}
where $\mathcal{D}^2|_\text{power-law}$ denotes the set of all network models that have a power-law DD in the asymptotic limit $k\to\infty$, and $\mathcal{D}^4|_\text{power-law}$ for power-law DDDD in the asymptotic limit $\eta \to \infty$.
We derive Eq.~\eqref{eq_main} in two steps,  proving its (i) {inclusion} and (ii) {strict inequality} as follows:

(i) $\mathcal{D}^2|_\text{power-law} \subseteq \mathcal{D}^4|_\text{power-law}$, i.e.,~\emph{every network with a power-law DD {also} has a power-law DDDD}. To prove this, we start by showing that given any power-law DD $f(k)\sim k^{-\alpha}$, the DDDD of the network can always be represented by two identical copies of the DD using the {copula theory}~\cite{copula_ds10}. Let $\mathcal{P}\left(\{k_i,k_j\}|i\leftrightarrow j \right)$ be the conditional joint probability of sequentially selecting two nodes $i$ and $j$ that have degrees $k_i,k_j\in [k_{\min},\infty)$, respectively, conditioned on $i$ and $j$ being connected. This implies that $\mathcal{P}\left(\{k_i,k_j\}|i\leftrightarrow j \right)$ can be understood as half the probability of selecting a link (from all links) that connects two nodes of degrees $k_i$ and $k_j$ (half because of the symmetry between $k_i$ and $k_j$). Its marginal distribution on $k_i$ can be obtained by integrating over $k_j$. This marginal distribution can be understood as the probability of selecting a link that is connected to a node of degree $k_i$. As a result, the marginal distribution is proportional to the DD, $f(k_i)$, times $k_i$, since a node with degree $k_i$ has a $k_i$ times greater chance of being selected. When $f(k_i)\propto k_i^{-\alpha}$, one has $\int dk_j  \mathcal{P}\left(\{k_i,k_j\}|i\leftrightarrow j \right)\propto k_i f(k_i) \sim k_i^{-\alpha+1}$. Thus, by Sklar's theorem~\cite{sklar_s59}, one can reversely write
\begin{equation}
\label{eq_copula}
\mathcal{P}\left(\{k_i,k_j\}|i\leftrightarrow j \right)\sim k_i^{-\alpha+1} k_j^{-\alpha+1} c(\frac{k_i^{-\alpha+2}}{k_{\min}^{-\alpha+2}},\frac{k_j^{-\alpha+2}}{k_{\min}^{-\alpha+2}}), 
\end{equation}
in terms of the marginal distributions $k_i^{-\alpha+1}$ and $k_j^{-\alpha+1}$ multiplied by a copula density $c(x, y)\ge 0$ defined on $(x,y)\in [0,1]^2$. 
Following Ref.~\cite{degree-degree-distance_zms20} (also see SI), the DDDD is given by
\begin{equation}
\label{eq_dddd_recipe}
g(\eta)=\int_{k_{\min}}^{\infty}2 \mathcal{P}\left(\{k_i,\eta k_i\}|i\leftrightarrow j \right) k_i dk_i.
\end{equation}
Inserting Eq.~\eqref{eq_copula} into Eq.~\eqref{eq_dddd_recipe} yields the DDDD,
\begin{equation*}
	g(\eta)\sim \int_{k_{\min}}^{\infty} \hspace{-2mm} k_i^{-2\alpha+2} \eta^{-\alpha+1} \left(c(\frac{k_i^{-\alpha+2}}{k_{\min}^{-\alpha+2}},0)+ O(\eta^{-\alpha+2})\right)k_i dk_i \quad 
\end{equation*}
for large $\eta$, which always contains a positive power-law leading term $\sim\eta^{-\alpha+1}=\eta^{-\beta}$ when $\alpha>2$, provided that the marginal copula density $c(x,0)>0$ on $x\in (0,1)$, satisfied as long as the copula density is smooth. We thus confirm the relationship $\beta=\alpha-1$ as empirically observed~\cite{degree-degree-distance_zms20}.

(ii)  $\mathcal{D}^2|_\text{power-law} \neq \mathcal{D}^4|_\text{power-law}$, i.e.,~\emph{there are networks that do not have a power-law DD but exhibit a power-law DDDD}.
To prove this, it suffices to provide a counterexample with a power-law DDDD but not a power-law DD. Here, we present {two} network models that serve as counterexamples. These models provide theoretical insights through two very distinct mechanisms (preferential attachment vs.~fitness), both having practical applications for modeling real-world networks.

\begin{figure*}[!t]
	\centering
    \includegraphics[width=12cm]{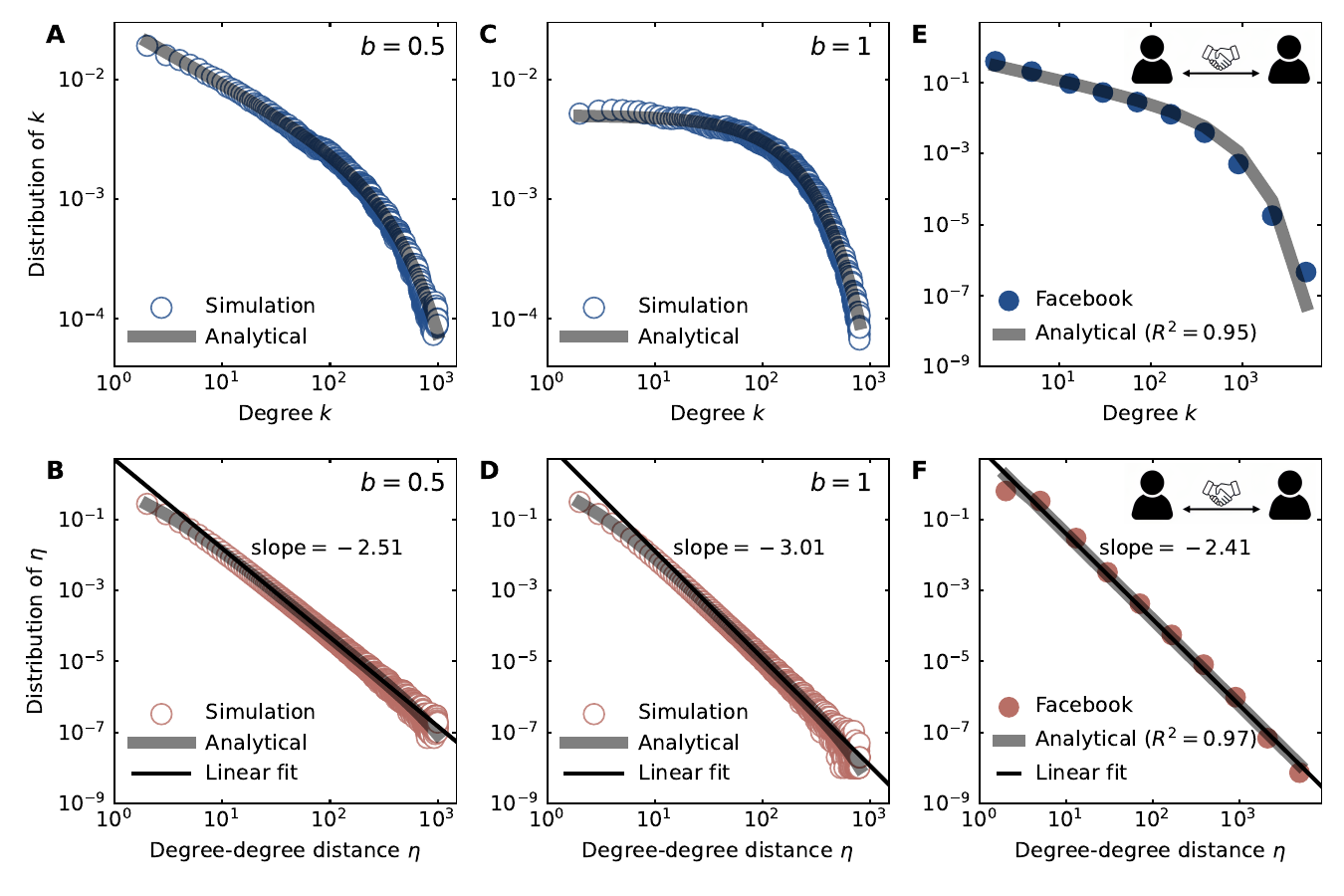}
	\caption{
		\label{fig_prefer}
		Distributions of (a,c)~$k$ and (b,d)~$\eta$ for the ``no-growth'' model, a network model of a fixed number of nodes, between which each internal link is added at each time step following the preferential attachment rule. The model simulation results (circle) are given by the average of $10^2$ runs on $N=10^4$ nodes and $T=10^6$ links. (e,f) Distributions of (e)~$k$ and (f)~$\eta$ of a subnetwork of Facebook, consisting of $N\sim10^6$ nodes (users) and $T\sim10^8$ links (user followerships). Both $k$ and $\eta$ are fitted by the no-growth model with $b=0.41$ under the same $N$ and $T$, revealing that the Facebook network exhibits a power-law distribution of $\eta$ rather than $k$. 
		\hfill\hfill}
\end{figure*}

\subsection*{Preferential attachment of internal links}
Our first model is a ``no-growth'' preferential attachment model, where the number of nodes $N$ is fixed, and only internal links (connections between existing nodes) are added to the network during its evolution. At every time step $t$, the model randomly selects two nodes $i$ and $j$, creating a link between them. The probability of selecting these two nodes is $\propto \left(k_i+b\right)\left(k_j+b\right)$, where $b$ is a small constant. The network, initially empty, acquires $T$ links after $T$ time steps. Although we allow multilinks so that $T$ can technically be larger than $N\left(N-1\right)/2$, our focus is on the limiting regime $T\sim N^{1+\delta}$ where $\delta$ is a small nonnegative number.  In this regime, we assume that multilinks become negligible as $N\to \infty$. Similar to the Barabási--Albert (BA) model~\cite{barabasi-albert_ba99}, the no-growth model exhibits the ``rich-gets-richer'' social phenomenon due to the linear dependence of the preferential attachment probability on the node degree. However, unlike the BA model, it does not capture the ``first-mover advantage'' phenomenon that is typically observed in marketing and business~\cite{netw-sci}.

To solve the model, we denote its DD by $f(T,k)$, the probability that a node has degree $k$ at time step $T$. The Markovian rate equation for $f(T,k)$ in the continuum limit can be written as (see SI)
\begin{equation}
\label{eq_prefer_continuous}
T'\frac{\partial f}{\partial T'}=-\frac{\partial }{\partial k'}\left(k' f\right),
\end{equation}
where $T'=T+bN/2$ and $k'=k+b$, both converging to $T$ and $k$ as $b\to0$. 
When $b$ is small, we find that the solution of Eq.~\eqref{eq_prefer_continuous} approximately follows a Gamma distribution for $k\in[0,\infty)$,
\begin{equation}
\label{eq_prefer_continuous_gamma}
f(k;A,B(T))=e^{-{B(T)}{k}} k^{A-1} B(T)^{A} /\Gamma(A).
\end{equation}
The parameters 
$A=b$ and $B(T)=bN/\left(2T\right)$ are fixed by matching the boundary condition $f(T,0)$ at $k=0$ (see SI). When $b$ is small, both the numerical solution of the rate equation [Eq.~\eqref{eq_prefer_continuous}] and the approximate analytical solution [Eq.~\eqref{eq_prefer_continuous_gamma}]  accurately reproduce the model simulation result [Fig.~\ref{fig_prefer}(a,c)].

The DD of the no-growth model is far from being a power law: when $b\ge 1$, Eq.~\eqref{eq_prefer_continuous_gamma} exhibits no heavy tail but an exponential upper tail [Fig.~\ref{fig_prefer}(c)]; when $b< 1$, Eq.~\eqref{eq_prefer_continuous_gamma} might be interpreted as an almost flat power law plus a finite-size exponential cutoff [Fig.~\ref{fig_prefer}(a)], but a power-law exponent $1-b$ that is less than one is often considered nonphysical~\cite{cohen2003structural}. Our result differs from the previous finding that preferential attachment of internal links would yield a power-law DD with an exponent of $2+b/2$~\cite{power-law-expon_gcb13}. This is because the previous study took a different limit under the assumption of stationary growth, which is not applicable to the no-growth model.

Note that Eq.~\eqref{eq_prefer_continuous} admits dilation symmetries~\cite{conform-field-theor-3d} under two scale transformations $T'\to T'+ \epsilon T'$ and $k'\to k'+ \epsilon k'$, respectively. This suggests that the rate equation [Eq.~\eqref{eq_prefer_continuous}] is intrinsically scale-free. The observation that $f(T,k)$ is not power law is thus a result of spontaneous symmetry breaking: when the initial condition is set to a normalized Dirac function, $f(0,k)=\delta(k)$, it breaks the dilation symmetry in the solution of Eq.~\eqref{eq_prefer_continuous}.

To derive the DDDD, recall that the probability of selecting a link that is connected to a node of degree $k$ is proportional to $k f(k;A,B(T))$. When the DD follows a Gamma distribution [Eq.~\eqref{eq_prefer_continuous_gamma}], this probability is simplified to $f(k;A+1,B(T))$. Similar to Eq.~\eqref{eq_copula}, we can write 
\begin{equation*}
\mathcal{P}\left(\{k_i,k_j\}|i\leftrightarrow j \right) =     f(k_i;A+1,B(T)) f(k_j;A+1,B(T))
\end{equation*}
as the product of two marginal distributions, where now the copula density term disappears as $k_i$ and $k_j$ are independent from each other. The DDDD at time step $T$, denoted by $g(T,\eta)$, is thus given by [Eq.~\eqref{eq_dddd_recipe}]
\begin{eqnarray}
\label{eq_prefer_continuous_eta}
&&g(T,\eta)
=\int_{1}^{\infty}2f(\eta k;A+1,B(T))f(k;A+1,B(T))kdk\nonumber\\
&&={2 B(T) ^{2 A+2} \eta ^A E_{-2 A-1}(\left(\eta+1\right)B(T) )}/\Gamma (A+1)^2,
\end{eqnarray}
where $E_n(x)$ is the exponential integral function.
This is not yet a power law. When the network is dense, however, expanding Eq.~\eqref{eq_prefer_continuous_eta} around {$T/N\to\infty$} yields
\begin{eqnarray*}
\label{eq_prefer_continuous_eta_exponent}
g(T,\eta)&\simeq&
\frac{2 \eta ^b (\eta +1)^{-2 (b+1)} \Gamma (2 (b+1))}{\Gamma (b+1)^2}+O(T^{-2-2b})\nonumber\\
&\sim&\eta^{-2-b}
\qquad \text{for large }T/N\text{, large }\eta.
\end{eqnarray*}
In contrast to DD, the DDDD of the no-growth model exhibits a strong power law (when $b$ is small). The analytical solution [Eq.~\eqref{eq_prefer_continuous_eta}] matches the model simulation result [Fig.~\ref{fig_prefer}(b,d)], with a power-law exponent of $2+b$ as expected.

We tested the empirical validity of the no-growth model by applying it to the social media network, \emph{Facebook}~\cite{facebook_gkbm10}. We find that both its DD and DDDD are accurately reproduced by the no-growth model under the same 
$N$ and $T$ [Fig.~\ref{fig_prefer}(e,f)]. Remarkably, the accuracy of the model is maintained over the full range of $k$ and $\eta$ by fitting only one parameter, $b$. 
Our result suggests that the no-growth model better captures the underlying mechanism of the Facebook network than the BA model---indeed, while preferential attachment 
is evident (as users with higher degrees are more attractive to others~\cite{prefer-attach_kbm13}), as a mature social platform, Facebook is unlikely to experience further growth in its number of users. Therefore, the no-growth model, which does not rely on network growth, is a better fit for the Facebook network.

Note that both the DD and DDDD of the no-growth model depend on the ratio $N/T$, not on $N$ or $T$ separately. This self-similarity allows us to estimate the properties of the entire network (e.g.,~Facebook) by analyzing a relatively small subnetwork, as long as the ratio $N/T$ remains the same. The proportionality between $N$ and $T$ is reminiscent of the self-similarity of the BA model, where $N/T$ also remains the same during the growth of the network. Yet, while the BA model only remains in the sparse regime ($N\sim T$), the no-growth model also extends to the dense regime ($N\ll T$). Indeed, the denser the network is, the closer $N/T$ is to zero, and hence the DDDD of the no-growth model is closer to a power law.

Another connection between the no-growth model and the BA model is that when taking $b\to0$ in the no-growth model, we find a DDDD of $\sim\eta^{-2}$, which is the same as the DDDD of the BA model (see SI). It seems that both models share the same scale-free property, characterized by the same power-law exponent of DDDD, which is \emph{solely} attributed to the preferential attachment mechanism. This is different from the power-law DD in the BA model, which is attributed to both preferential attachment and growth~\cite{netw-sci}. Indeed, a BA model without growth (``model B''~\cite{netw-sci}) cannot have a power-law DD.

\begin{figure*}[!t]
	\centering
    \includegraphics[width=12cm]{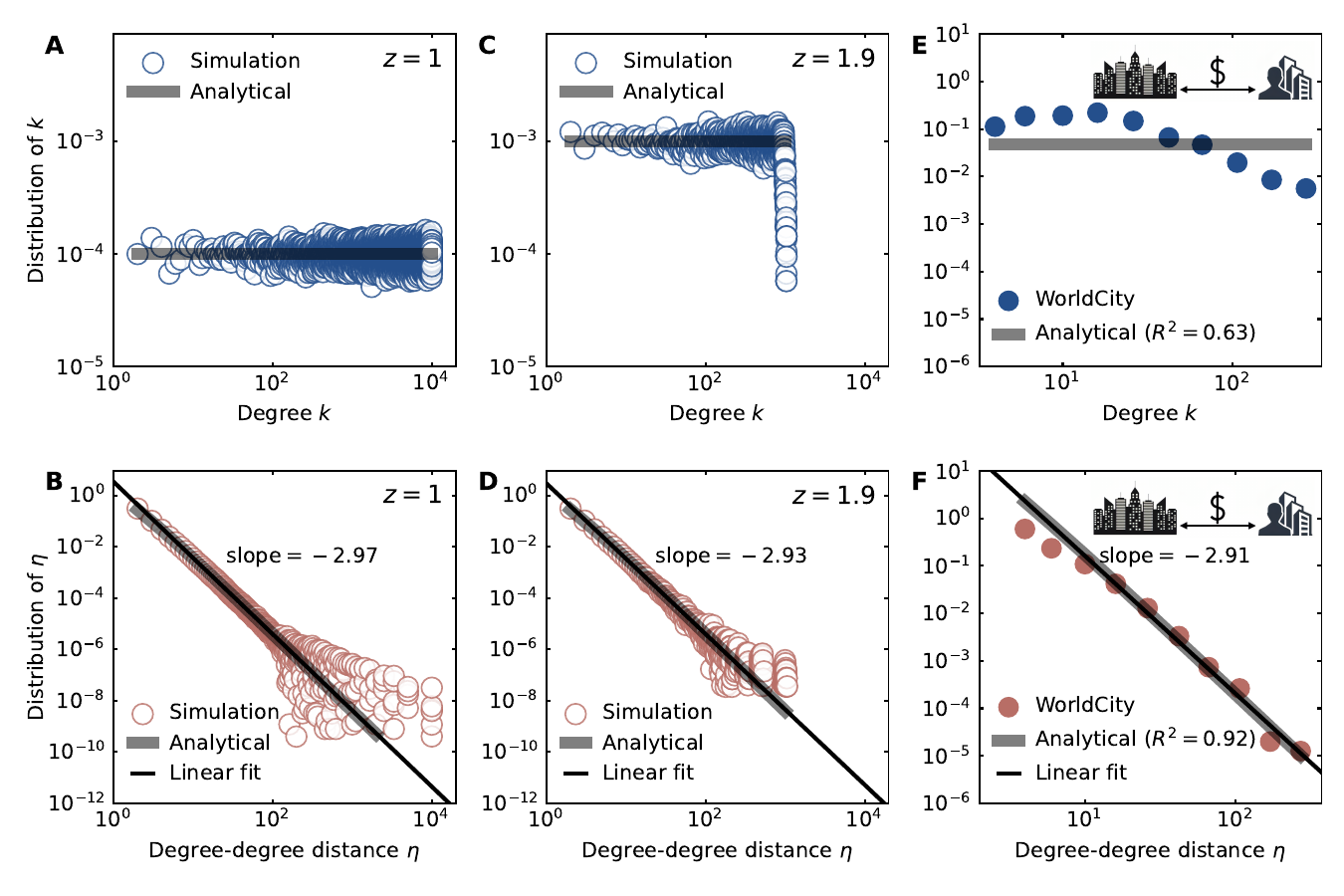}
	\caption{
		\label{fig_fitness}
		Distributions of (a,c)~$k$ and (b,d)~$\eta$ for the "hard-threshold" model, a network model that assigns a uniformly distributed fitness $\omega\in[0,\omega_{\max}=1]$ to each node, where two nodes are linked if and only if their fitness sum is larger than a threshold $z$. The model simulation results (circle) are given by the average of $10^2$ runs on $N=10^4$ nodes. (e,f) Distributions of (e)~$k$ and (f)~$\eta$ of the WorldCity network, an urban service network consisting of $N=415$ nodes (global cities and service firms) and $T=7,518$ links (demand-supply relations between cities and firms). Both $k$ and $\eta$ are fitted by the hard-threshold model (solid lines) with $z=1.95\omega_{\max}$ under the same $N$ and $T$, revealing that the WorldCity network exhibits a power-law distribution of $\eta$ rather than $k$.
		\hfill\hfill}
\end{figure*}

\subsection*{Fitness with threshold}
Our second model is defined by assigning a random fitness $\omega$ that follows a fitness distribution $\rho(\omega)$ to each node of the
network~\cite{bianconi-barabasi_bb01}. For every two nodes $i$ and $j$, let a link be drawn with probability $\sigma(\omega_i,\omega_j)$ that depends only on the fitness of the nodes, not their degrees~\cite{quench-fitness_ccdm02}. By choosing  $\sigma(\omega_i,\omega_j)= \theta(\omega_i+\omega_j-z)$, where $\theta(x)$ is the Heaviside step function, a "hard threshold" of fitness is introduced, such that a link will be deterministically drawn if and only if the sum of the fitnesses of the two nodes is larger than a {threshold} $z$. As a result, the network topology is only determined by  {quenched disorder} from the randomness of fitness, not by disorder of network evolution~\cite{quench-fitness_ccdm02}. This hard thresholding rule simplifies success or failure as a binary event and emphasizes the collaborative nature of success, which depends solely on the combined capabilities of both parties involved.

The hard-threshold model was first solved for $\rho(\omega)=e^{-\omega}$ (denoting an exponential fitness distribution), for which the DD is found to be 
$f(k)\sim k^{-2}$, with a power-law exponent of two~\cite{quench-fitness_scb04}. However, the DD is not power law in general for arbitrary $\rho(\omega)$. Here, we consider a uniform distribution $\rho(\omega)$ for $\omega\in[0,\omega_{\max}]$ and assume that $\omega_{\max}\le z\le2\omega_{\max}$. The conditional probability that a node has degree $k$ given fitness $\omega_i$ is given by
\begin{equation}
\label{eq_fitness_quenched}
\mathcal{P}(k|\omega_i)\simeq \binom{N}{k}p(\omega_i)^{k}\left(1-p(\omega_i)\right)^{N-k},
\end{equation}
where $p(\omega_i)=\max\{1-\left(z-\omega_i\right)/\omega_{\max},0\}$ is the probability that another selected node $j$ has a fitness $\omega_j$ larger than $z-\omega_i$ so that nodes $i$ and $j$ are linked. Approximating Eq.~\eqref{eq_fitness_quenched} by a Gaussian distribution and taking the large $N$ limit,
we find that the DD is given by  (see SI)
\begin{equation}
\label{DD_fitness_model}
f(k)=\int_0^{\omega_{\max}}
\hspace{-2mm}
\mathcal{P}(k|\omega_i)\rho(\omega_i) d\omega_i\simeq \frac{N^{-1}}{2-z/\omega_{\max}}+O(N^{-2}),
\end{equation} 
which is flat and independent of $k$ [Fig.~\ref{fig_fitness}(a,c)].

To derive the DDDD, 
let $\mathcal{P}\left(\{\omega_i,\omega_j\}|i\leftrightarrow j\right)$ be the conditional joint probability of sequentially selecting two nodes $i$ and $j$ of fitnesses $\omega_i$ and $\omega_j$, conditioned on $i$ and $j$ being connected. Using Bayes' rule, we have $\mathcal{P}\left(\{\omega_i,\omega_j\}|i\leftrightarrow j\right) = {\mathcal{P}\left(i\leftrightarrow j|\{\omega_i,\omega_j\}\right)~\cdot~\mathcal{P}\left(\{\omega_i,\omega_j\}\right)}~/~{\mathcal{P}\left(i\leftrightarrow j\right)}\simeq  \theta(\omega_i+\omega_j-z)\cdot \omega_{\max}^{-2} N^2 T^{-1}/2$.
Thus, by Eq.~\eqref{eq_dddd_recipe} we have (see SI)
\begin{eqnarray}
\label{eq_fitness_eta}
\hspace{-10mm}&&g(\eta)=
\frac{\omega_{\max}^{-2} N^2}{2T}
\int_{1}^{\infty}\hspace{-1mm}2kdk \int_{z-\omega_{\max}}^{\omega_{\max}}\hspace{-6mm}d\omega_i \int_{z-\omega_{i}}^{\omega_{\max}}\hspace{-4mm}d\omega_j 
\mathcal{P}(\eta k|\omega_i)\mathcal{P}(k|\omega_j)\nonumber\\
\hspace{-10mm}&&\simeq
\frac{ N^2 (2-z/\omega_{\max})^2\left(2 \eta +1\right) }{2 T\eta ^2 \left(\eta +1\right)^2}+O(N^1), 
\end{eqnarray}
which yields
\begin{equation*}
    g(\eta) \sim \eta^{-3},\qquad \text{for large }\eta
\end{equation*}
a strong power law that is independent of $z$ [Fig.~\ref{fig_fitness}(b,d)]. 

Applying the hard-threshold model to real-world networks, we find that the model is in good agreement with the \emph{WordCity} network~\cite{taylor2015world}, a network of global cities and service firms, which also exhibits an almost flat DD and a power-law DDDD of an exponent equal to $3$ [Fig.~\ref{fig_fitness}(e,f)]. Again, the agreement is maintained over the full range of $k$ and $\eta$ by fitting only one parameter, $z/\omega_{\max}$. This indicates that the hard-threshold model captures how business relationships are formed globally.  Our result highlights the \emph{threshold effect} in international trade~\cite{grossman2017matching}, that a business relationship is established only when the potential combined benefits of the two parties exceed a particular threshold. Furthermore, the use of the fitness model reflects the decision-making process of global business, which is arguably more based on intrinsic values (fitness) rather than random evolution (e.g.,~preferential attachment).

\begin{figure}[t!]
	\centering
	\includegraphics[width=\linewidth]{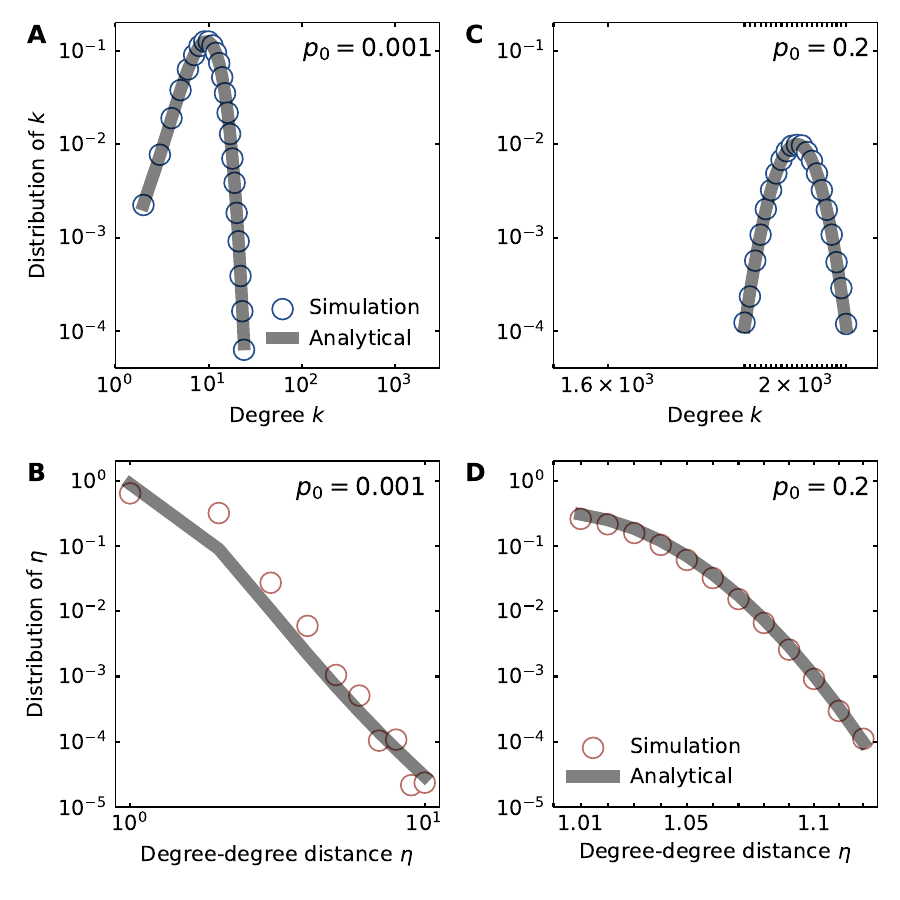}
	\vspace{-7mm}
	\caption{
		\label{fig_er_dddd}
		Distributions of (a,c)~$k$ and (b,d)~$\eta$ for the Erdős--Rényi model, given the linking probability $p_0$ fixed. The model simulation results (circle) are given by the average of $10^2$ runs on $N=10^4$ nodes.
		\hfill\hfill}
\end{figure}

\subsection*{The Erdős--Rényi model revisited}
For both the no-growth and the hard-threshold models, we have shown that the DDDD is power law even if the DD is not. Now we give a case where neither DD nor DDDD is power law. Considering the Erdős--Rényi (ER) model~\cite{erdos-renyi_er59}, we know that its DD $f(k)\simeq \binom{N}{k}p_0^{k}\left(1-p_0\right)^{N-k}$ is not a power law but approximately a Gaussian distribution when taking the large $N$ limit while fixing $p_0$---the probability to link two random nodes. We find that the DDDD of the ER model (see SI), 
\begin{eqnarray}
\label{eq_er_eta}
g(\eta)
\simeq
\left[\frac{2 N p_0}{\pi\left(1-p_0\right)}\right]^{1/2}
\frac{\eta \left(\eta +1\right)^3}{\left(\eta ^2+1\right)^{7/2}}
e^{-\frac{N p_0 \left(\eta -1\right)^2}{2 \left(1-p_0\right) \left(\eta ^2+1\right)}},
\end{eqnarray}
is not power law when fixing $p_0$. This is since $\eta \le N$ by definition [Eq.~\eqref{eq_eta}],  which sets up an upper bound on $\eta$. As a result, the exponential term in Eq.~\eqref{eq_er_eta} can never vanish for large $\eta$ (Fig.~\ref{fig_er_dddd}). Note that the ER model can also be derived from both of our models, either by taking $b\to\infty$ and letting $T=N^2p_0/2$ in the no-growth
model, or by completely softening the threshold so that $\sigma(\omega_i,\omega_j)\to p_0$ becomes independent of $\omega_i$ and $\omega_j$ in the hard-threshold model. This suggests that the DDDD is power law only in specific regimes, depending on the range of the parameters and/or the order of taking their separate limits. Hence, there could be multiple regimes, characterized by different power laws or no power law of $\eta$ at all. This leads to an open question to identify the \emph{crossover}~\cite{ma2020role, crossover-sir-peak_mmb22} of power laws between these regimes.

\section*{Discussion and Conclusions}

From a data science perspective, notions on heavy-tail distributions seem to be more relevant than power-law distributions for pure quantitative purposes such as statistical analysis and data processing~\cite{power-law-heavy-tail_sp12}.
Knowing the heaviness of distribution tails can certainly help estimating the frequency of extreme events, beyond which fitting a power law may not provide much additional interpretation. Yet, there is an equally important question to \emph{how}---and that is \emph{why}. Understanding the underlying mechanisms that drive network generation is just as important as knowing how to analyze the data. Simply labeling a distribution as heavy-tailed does not reveal these mechanisms, and this is where the scale-free property of networks comes in. We show that the DDDD is a valuable tool for uncovering such mechanisms. Indeed, without considering the power-law DDDD, we would not have been able to infer the scale-free mechanisms of a variety of empirical complex systems, including the Facebook and the WorldCity examples.

We suggest that similar to DD, the power-law exponent of the DDDD could also be incorporated into network theories, such as percolation or dynamical systems on networks, particularly when the DD of the network is not power law. This could expand the applicability of scale-free theories to a broader scope of networks that were previously not considered scale-free. 

Furthermore, Eq.~\eqref{eq_main} establishes a strictly inclusive order between the power laws of DD and DDDD. This raises the question of whether a ``hierarchy'' of power laws can be established for other known or unknown network metrics, providing new insights into the origins of scale-freeness.

In conclusion, we focus on the relationship between the degree distribution (DD) and the degree--degree distance distribution (DDDD) of network models, demonstrating that power-law DDDD is more universal than power-law DD. This prompts us to consider the DDDD as a more suitable indicator of scale-free networks and reach a deeper understanding of their universality and underlying mechanisms.

\section*{Acknowledgements}

This work started from a discussion with H.~E.~Stanley. We thank H.~E.~Stanley, P.~Holme, B.~Chen, and J.~Ma for fruitful discussion. X.M. was supported by the NetSeed: Seedling Research Award of Northeastern University. B.Z. was supported by the Startup Foundation for Introducing Talent of NUIST and the Qinglan Project of Jiangsu Universities, and was supported by National Natural Science Foundation of China with Grant Number 72074101.

\section*{Data Availability}

The Facebook and WorldCity data are publicly available at \url{https://athinagroup.eng.uci.edu} and \url{http://vlado.fmf.uni-lj.si/pub/networks/data/}

\section*{Author Contributions}

X.M.~and B.Z.~conceived the research, conducted the analysis, and wrote the manuscript. 

\section*{Competing Interests}

The authors declare no competing interests.

\section*{Correspondence}

Correspondence and requests for materials should be addressed to B.Z.\newline (email: binzhou@mail.ustc.edu.cn).

\bibliography{DDDDPowerLaw}

\begin{thebibliography}{10}
\urlstyle{rm}
\expandafter\ifx\csname url\endcsname\relax
  \def\url#1{\texttt{#1}}\fi
\expandafter\ifx\csname urlprefix\endcsname\relax\def\urlprefix{URL }\fi
\expandafter\ifx\csname doiprefix\endcsname\relax\def\doiprefix{DOI: }\fi
\providecommand{\bibinfo}[2]{#2}
\providecommand{\eprint}[2][]{\url{#2}}

\bibitem{netw-introd}
\bibinfo{author}{Newman, M. E.~J.}
\newblock \emph{\bibinfo{title}{Networks: {{An Introduction}}}}
  (\bibinfo{publisher}{{Oxford University Press}}, \bibinfo{address}{{New
  York}}, \bibinfo{year}{2010}), \bibinfo{edition}{1st} edn.

\bibitem{stat-of-netw_ab02}
\bibinfo{author}{Albert, R.} \& \bibinfo{author}{Barab{\'a}si, A.-L.}
\newblock \bibinfo{journal}{\bibinfo{title}{Statistical mechanics of complex
  networks}}.
\newblock {\emph{\JournalTitle{Rev. Mod. Phys.}}}
  \textbf{\bibinfo{volume}{74}}, \bibinfo{pages}{47--97}
  (\bibinfo{year}{2002}).

\bibitem{complex-crit}
\bibinfo{author}{Christensen, K.} \& \bibinfo{author}{Moloney, N.~R.}
\newblock \emph{\bibinfo{title}{Complexity and {{Criticality}}}},
  vol.~\bibinfo{volume}{1} of \emph{\bibinfo{series}{Imperial {{College Press
  Advanced Physics Texts}}}} (\bibinfo{publisher}{{Imperial College Press}},
  \bibinfo{address}{{London}}, \bibinfo{year}{2005}), \bibinfo{edition}{1st}
  edn.

\bibitem{barabasi-albert_ba99}
\bibinfo{author}{Barab{\'a}si, A.-L.} \& \bibinfo{author}{Albert, R.}
\newblock \bibinfo{journal}{\bibinfo{title}{Emergence of scaling in random
  networks}}.
\newblock {\emph{\JournalTitle{Science}}} \textbf{\bibinfo{volume}{286}},
  \bibinfo{pages}{509--512} (\bibinfo{year}{1999}).

\bibitem{stat-field-theor-1}
\bibinfo{author}{Itzykson, C.} \& \bibinfo{author}{Drouffe, J.-M.}
\newblock \emph{\bibinfo{title}{Statistical {{Field Theory}}: {{Volume}} 1,
  {{From Brownian Motion}} to {{Renormalization}} and {{Lattice Gauge
  Theory}}}} (\bibinfo{publisher}{{Cambridge University Press}},
  \bibinfo{address}{{New York}}, \bibinfo{year}{1989}), \bibinfo{edition}{1st}
  edn.

\bibitem{stat-field-theor-2}
\bibinfo{author}{Itzykson, C.} \& \bibinfo{author}{Drouffe, J.-M.}
\newblock \emph{\bibinfo{title}{Statistical {{Field Theory}}: {{Volume}} 2,
  {{Strong Coupling}}, {{Monte Carlo Methods}}, {{Conformal Field Theory}} and
  {{Random Systems}}}} (\bibinfo{publisher}{{Cambridge University Press}},
  \bibinfo{address}{{New York}}, \bibinfo{year}{1989}), \bibinfo{edition}{1st}
  edn.

\bibitem{scale-free_t05}
\bibinfo{author}{Tanaka, R.}
\newblock \bibinfo{journal}{\bibinfo{title}{Scale-{{Rich Metabolic
  Networks}}}}.
\newblock {\emph{\JournalTitle{Phys. Rev. Lett.}}}
  \textbf{\bibinfo{volume}{94}}, \bibinfo{pages}{168101}
  (\bibinfo{year}{2005}).

\bibitem{comment-to-barabasi-albert_ah00}
\bibinfo{author}{Adamic, L.~A.} \& \bibinfo{author}{Huberman, B.~A.}
\newblock \bibinfo{journal}{\bibinfo{title}{Power-law distribution of the
  {{World Wide Web}}}}.
\newblock {\emph{\JournalTitle{Science}}} \textbf{\bibinfo{volume}{287}},
  \bibinfo{pages}{2115--2115} (\bibinfo{year}{2000}).

\bibitem{krapivsky2001degree}
\bibinfo{author}{Krapivsky, P.~L.}, \bibinfo{author}{Rodgers, G.~J.} \&
  \bibinfo{author}{Redner, S.}
\newblock \bibinfo{journal}{\bibinfo{title}{Degree distributions of growing
  networks}}.
\newblock {\emph{\JournalTitle{Phys. Rev. Lett.}}}
  \textbf{\bibinfo{volume}{86}}, \bibinfo{pages}{5401} (\bibinfo{year}{2001}).

\bibitem{netw-self-similar_shm05}
\bibinfo{author}{Song, C.}, \bibinfo{author}{Havlin, S.} \&
  \bibinfo{author}{Makse, H.~A.}
\newblock \bibinfo{journal}{\bibinfo{title}{Self-similarity of complex
  networks}}.
\newblock {\emph{\JournalTitle{Nature}}} \textbf{\bibinfo{volume}{433}},
  \bibinfo{pages}{392--395} (\bibinfo{year}{2005}).

\bibitem{evans2005scale}
\bibinfo{author}{Evans, T.~S.} \& \bibinfo{author}{Saram{\"a}ki, J.}
\newblock \bibinfo{journal}{\bibinfo{title}{Scale-free networks from
  self-organization}}.
\newblock {\emph{\JournalTitle{Phys. Rev. E}}} \textbf{\bibinfo{volume}{72}},
  \bibinfo{pages}{026138} (\bibinfo{year}{2005}).

\bibitem{crit-phenom-netw_dgm08}
\bibinfo{author}{Dorogovtsev, S.~N.}, \bibinfo{author}{Goltsev, A.~V.} \&
  \bibinfo{author}{Mendes, J. F.~F.}
\newblock \bibinfo{journal}{\bibinfo{title}{Critical phenomena in complex
  networks}}.
\newblock {\emph{\JournalTitle{Rev. Mod. Phys.}}}
  \textbf{\bibinfo{volume}{80}}, \bibinfo{pages}{1275--1335}
  (\bibinfo{year}{2008}).

\bibitem{szell2010multirelational}
\bibinfo{author}{Szell, M.}, \bibinfo{author}{Lambiotte, R.} \&
  \bibinfo{author}{Thurner, S.}
\newblock \bibinfo{journal}{\bibinfo{title}{Multirelational organization of
  large-scale social networks in an online world}}.
\newblock {\emph{\JournalTitle{Proc. Natl. Acad. Sci.}}}
  \textbf{\bibinfo{volume}{107}}, \bibinfo{pages}{13636--13641}
  (\bibinfo{year}{2010}).

\bibitem{lambiotte2019networks}
\bibinfo{author}{Lambiotte, R.}, \bibinfo{author}{Rosvall, M.} \&
  \bibinfo{author}{Scholtes, I.}
\newblock \bibinfo{journal}{\bibinfo{title}{From networks to optimal
  higher-order models of complex systems}}.
\newblock {\emph{\JournalTitle{Nat. Phys.}}} \textbf{\bibinfo{volume}{15}},
  \bibinfo{pages}{313--320} (\bibinfo{year}{2019}).

\bibitem{power-law-outage_nsz20}
\bibinfo{author}{Nesti, T.}, \bibinfo{author}{Sloothaak, F.} \&
  \bibinfo{author}{Zwart, B.}
\newblock \bibinfo{journal}{\bibinfo{title}{Emergence of {{Scale-Free Blackout
  Sizes}} in {{Power Grids}}}}.
\newblock {\emph{\JournalTitle{Phys. Rev. Lett.}}}
  \textbf{\bibinfo{volume}{125}}, \bibinfo{pages}{058301}
  (\bibinfo{year}{2020}).

\bibitem{netw-percolation_ceah00}
\bibinfo{author}{Cohen, R.}, \bibinfo{author}{Erez, K.},
  \bibinfo{author}{{ben-Avraham}, D.} \& \bibinfo{author}{Havlin, S.}
\newblock \bibinfo{journal}{\bibinfo{title}{Resilience of the {{Internet}} to
  {{Random Breakdowns}}}}.
\newblock {\emph{\JournalTitle{Phys. Rev. Lett.}}}
  \textbf{\bibinfo{volume}{85}}, \bibinfo{pages}{4626--4628}
  (\bibinfo{year}{2000}).

\bibitem{netw-resil_gbb16}
\bibinfo{author}{Gao, J.}, \bibinfo{author}{Barzel, B.} \&
  \bibinfo{author}{Barab{\'a}si, A.-L.}
\newblock \bibinfo{journal}{\bibinfo{title}{Universal resilience patterns in
  complex networks}}.
\newblock {\emph{\JournalTitle{Nature}}} \textbf{\bibinfo{volume}{530}},
  \bibinfo{pages}{307--312} (\bibinfo{year}{2016}).

\bibitem{finite-size_scmrsbc21}
\bibinfo{author}{Serafino, M.} \emph{et~al.}
\newblock \bibinfo{journal}{\bibinfo{title}{True scale-free networks hidden by
  finite size effects}}.
\newblock {\emph{\JournalTitle{Proc. Natl. Acad. Sci.}}}
  \textbf{\bibinfo{volume}{118}}, \bibinfo{pages}{e2013825118}
  (\bibinfo{year}{2021}).

\bibitem{goh2002classification}
\bibinfo{author}{Goh, K.-I.}, \bibinfo{author}{Oh, E.}, \bibinfo{author}{Jeong,
  H.}, \bibinfo{author}{Kahng, B.} \& \bibinfo{author}{Kim, D.}
\newblock \bibinfo{journal}{\bibinfo{title}{Classification of scale-free
  networks}}.
\newblock {\emph{\JournalTitle{Proc. Natl. Acad. Sci.}}}
  \textbf{\bibinfo{volume}{99}}, \bibinfo{pages}{12583--12588}
  (\bibinfo{year}{2002}).

\bibitem{protein-interact_klswpsbkkhcvb19}
\bibinfo{author}{Kov{\'a}cs, I.~A.} \emph{et~al.}
\newblock \bibinfo{journal}{\bibinfo{title}{Network-based prediction of protein
  interactions}}.
\newblock {\emph{\JournalTitle{Nat. Commun.}}} \textbf{\bibinfo{volume}{10}},
  \bibinfo{pages}{1240} (\bibinfo{year}{2019}).

\bibitem{garcia2013social}
\bibinfo{author}{Garcia, D.}, \bibinfo{author}{Mavrodiev, P.} \&
  \bibinfo{author}{Schweitzer, F.}
\newblock \bibinfo{title}{Social resilience in online communities: The autopsy
  of friendster}.
\newblock In \emph{\bibinfo{booktitle}{Proceedings of the first ACM conference
  on Online social networks}}, \bibinfo{pages}{39--50} (\bibinfo{year}{2013}).

\bibitem{myers2014information}
\bibinfo{author}{Myers, S.~A.}, \bibinfo{author}{Sharma, A.},
  \bibinfo{author}{Gupta, P.} \& \bibinfo{author}{Lin, J.}
\newblock \bibinfo{title}{Information network or social network? the structure
  of the twitter follow graph}.
\newblock In \emph{\bibinfo{booktitle}{Proceedings of the 23rd International
  Conference on World Wide Web}}, \bibinfo{pages}{493--498}
  (\bibinfo{year}{2014}).

\bibitem{goh2001universal}
\bibinfo{author}{Goh, K.-I.}, \bibinfo{author}{Kahng, B.} \&
  \bibinfo{author}{Kim, D.}
\newblock \bibinfo{journal}{\bibinfo{title}{Universal behavior of load
  distribution in scale-free networks}}.
\newblock {\emph{\JournalTitle{Phys. Rev. Lett.}}}
  \textbf{\bibinfo{volume}{87}}, \bibinfo{pages}{278701}
  (\bibinfo{year}{2001}).

\bibitem{chen2002origin}
\bibinfo{author}{Chen, Q.}, \bibinfo{author}{Chang, H.},
  \bibinfo{author}{Govindan, R.} \& \bibinfo{author}{Jamin, S.}
\newblock \bibinfo{title}{The origin of power laws in internet topologies
  revisited}.
\newblock In \emph{\bibinfo{booktitle}{Proceedings. twenty-first annual joint
  conference of the ieee computer and communications societies}},
  vol.~\bibinfo{volume}{2}, \bibinfo{pages}{608--617}
  (\bibinfo{organization}{IEEE}, \bibinfo{year}{2002}).

\bibitem{power-law-is-rare_bc19}
\bibinfo{author}{Broido, A.~D.} \& \bibinfo{author}{Clauset, A.}
\newblock \bibinfo{journal}{\bibinfo{title}{Scale-free networks are rare}}.
\newblock {\emph{\JournalTitle{Nat. Commun.}}} \textbf{\bibinfo{volume}{10}},
  \bibinfo{pages}{1017} (\bibinfo{year}{2019}).

\bibitem{rare-and-everywhere_h19}
\bibinfo{author}{Holme, P.}
\newblock \bibinfo{journal}{\bibinfo{title}{Rare and everywhere:
  {{Perspectives}} on scale-free networks}}.
\newblock {\emph{\JournalTitle{Nat. Commun.}}} \textbf{\bibinfo{volume}{10}},
  \bibinfo{pages}{1016} (\bibinfo{year}{2019}).

\bibitem{voitalov2019scale}
\bibinfo{author}{Voitalov, I.}, \bibinfo{author}{van~der Hoorn, P.},
  \bibinfo{author}{van~der Hofstad, R.} \& \bibinfo{author}{Krioukov, D.}
\newblock \bibinfo{journal}{\bibinfo{title}{Scale-free networks well done}}.
\newblock {\emph{\JournalTitle{Physical Review Research}}}
  \textbf{\bibinfo{volume}{1}}, \bibinfo{pages}{033034} (\bibinfo{year}{2019}).

\bibitem{artico2020rare}
\bibinfo{author}{Artico, I.}, \bibinfo{author}{Smolyarenko, I.},
  \bibinfo{author}{Vinciotti, V.} \& \bibinfo{author}{Wit, E.}
\newblock \bibinfo{journal}{\bibinfo{title}{How rare are power-law networks
  really?}}
\newblock {\emph{\JournalTitle{Proceedings of the Royal Society A}}}
  \textbf{\bibinfo{volume}{476}}, \bibinfo{pages}{20190742}
  (\bibinfo{year}{2020}).

\bibitem{power-law-fit-range_csn09}
\bibinfo{author}{Clauset, A.}, \bibinfo{author}{Shalizi, C.~R.} \&
  \bibinfo{author}{Newman, M. E.~J.}
\newblock \bibinfo{journal}{\bibinfo{title}{Power-{{Law Distributions}} in
  {{Empirical Data}}}}.
\newblock {\emph{\JournalTitle{SIAM Rev.}}} \textbf{\bibinfo{volume}{51}},
  \bibinfo{pages}{661--703} (\bibinfo{year}{2009}).

\bibitem{degree-degree-distance_zms20}
\bibinfo{author}{Zhou, B.}, \bibinfo{author}{Meng, X.} \&
  \bibinfo{author}{Stanley, H.~E.}
\newblock \bibinfo{journal}{\bibinfo{title}{Power-law distribution of
  degree\textendash degree distance: {{A}} better representation of the
  scale-free property of complex networks}}.
\newblock {\emph{\JournalTitle{Proc. Natl. Acad. Sci.}}}
  \textbf{\bibinfo{volume}{117}}, \bibinfo{pages}{14812--14818}
  (\bibinfo{year}{2020}).

\bibitem{farzam2020degree}
\bibinfo{author}{Farzam, A.}, \bibinfo{author}{Samal, A.} \&
  \bibinfo{author}{Jost, J.}
\newblock \bibinfo{journal}{\bibinfo{title}{Degree difference: a simple measure
  to characterize structural heterogeneity in complex networks}}.
\newblock {\emph{\JournalTitle{Sci. Rep.}}} \textbf{\bibinfo{volume}{10}},
  \bibinfo{pages}{1--12} (\bibinfo{year}{2020}).

\bibitem{wang2021self}
\bibinfo{author}{Wang, B.}, \bibinfo{author}{Zhu, J.} \& \bibinfo{author}{Wei,
  D.}
\newblock \bibinfo{journal}{\bibinfo{title}{The self-similarity of complex
  networks: From the view of degree--degree distance}}.
\newblock {\emph{\JournalTitle{Mod. Phys. Lett. B}}}
  \textbf{\bibinfo{volume}{35}}, \bibinfo{pages}{2150331}
  (\bibinfo{year}{2021}).

\bibitem{maren20212}
\bibinfo{author}{Maren, A.~J.}
\newblock \bibinfo{journal}{\bibinfo{title}{The 2-d cluster variation method:
  Topography illustrations and their enthalpy parameter correlations}}.
\newblock {\emph{\JournalTitle{Entropy}}} \textbf{\bibinfo{volume}{23}},
  \bibinfo{pages}{319} (\bibinfo{year}{2021}).

\bibitem{log-degree_ec22}
\bibinfo{author}{Evans, T.~S.} \& \bibinfo{author}{Chen, B.}
\newblock \bibinfo{journal}{\bibinfo{title}{Linking the network centrality
  measures closeness and degree}}.
\newblock {\emph{\JournalTitle{Commun. Phys.}}} \textbf{\bibinfo{volume}{5}},
  \bibinfo{pages}{1--11} (\bibinfo{year}{2022}).

\bibitem{serafino2021true}
\bibinfo{author}{Serafino, M.} \emph{et~al.}
\newblock \bibinfo{journal}{\bibinfo{title}{True scale-free networks hidden by
  finite size effects}}.
\newblock {\emph{\JournalTitle{Proc. Natl. Acad. Sci.}}}
  \textbf{\bibinfo{volume}{118}}, \bibinfo{pages}{e2013825118}
  (\bibinfo{year}{2021}).

\bibitem{netw-mech_lb21}
\bibinfo{author}{Langendorf, R.~E.} \& \bibinfo{author}{Burgess, M.~G.}
\newblock \bibinfo{journal}{\bibinfo{title}{Empirically {{Classifying Network
  Mechanisms}}}}.
\newblock {\emph{\JournalTitle{Sci Rep}}} \textbf{\bibinfo{volume}{11}},
  \bibinfo{pages}{20501} (\bibinfo{year}{2021}).

\bibitem{quench-fitness_ccdm02}
\bibinfo{author}{Caldarelli, G.}, \bibinfo{author}{Capocci, A.},
  \bibinfo{author}{De~Los~Rios, P.} \& \bibinfo{author}{Mu{\~n}oz, M.~A.}
\newblock \bibinfo{journal}{\bibinfo{title}{Scale-{{Free Networks}} from
  {{Varying Vertex Intrinsic Fitness}}}}.
\newblock {\emph{\JournalTitle{Phys. Rev. Lett.}}}
  \textbf{\bibinfo{volume}{89}}, \bibinfo{pages}{258702}
  (\bibinfo{year}{2002}).

\bibitem{copula_ds10}
\bibinfo{author}{Durante, F.} \& \bibinfo{author}{Sempi, C.}
\newblock \bibinfo{title}{Copula {{Theory}}: {{An Introduction}}}.
\newblock In \bibinfo{editor}{Jaworski, P.}, \bibinfo{editor}{Durante, F.},
  \bibinfo{editor}{H{\"a}rdle, W.~K.} \& \bibinfo{editor}{Rychlik, T.} (eds.)
  \emph{\bibinfo{booktitle}{Copula {{Theory}} and {{Its Applications}}}}, vol.
  \bibinfo{volume}{198} of \emph{\bibinfo{series}{Lecture {{Notes}} in
  {{Statistics}}}}, \bibinfo{pages}{3--31} (\bibinfo{publisher}{{Springer}},
  \bibinfo{address}{{Berlin, Heidelberg}}, \bibinfo{year}{2010}).

\bibitem{sklar_s59}
\bibinfo{author}{Sklar, A.}
\newblock \bibinfo{journal}{\bibinfo{title}{Fonctions de r\'epartition \`a n
  dimensions et leurs marges}}.
\newblock {\emph{\JournalTitle{Publ. Inst. Stat. Univ. Paris}}}
  \textbf{\bibinfo{volume}{8}}, \bibinfo{pages}{229--231}
  (\bibinfo{year}{1959}).

\bibitem{netw-sci}
\bibinfo{author}{Barab{\'a}si, A.-L.}
\newblock \emph{\bibinfo{title}{Network {{Science}}}}
  (\bibinfo{publisher}{{Cambridge University Press}},
  \bibinfo{address}{{Boston}}, \bibinfo{year}{2016}), \bibinfo{edition}{1st}
  edn.

\bibitem{cohen2003structural}
\bibinfo{author}{Cohen, R.}, \bibinfo{author}{Havlin, S.} \&
  \bibinfo{author}{{ben-Avraham}, D.}
\newblock \bibinfo{title}{Structural properties of scale free networks}.
\newblock In \bibinfo{editor}{Bornholdt, S.} \& \bibinfo{editor}{Schuster,
  H.~G.} (eds.) \emph{\bibinfo{booktitle}{Handbook of {{Graphs}} and
  {{Networks}}: {{From}} the {{Genome}} to the {{Internet}}}},
  \bibinfo{pages}{85--110} (\bibinfo{publisher}{{Wiley-VCH}},
  \bibinfo{address}{{Berlin, Germany}}, \bibinfo{year}{2003}),
  \bibinfo{edition}{1st} edn.

\bibitem{power-law-expon_gcb13}
\bibinfo{author}{Ghoshal, G.}, \bibinfo{author}{Chi, L.} \&
  \bibinfo{author}{Barab{\'a}si, A.-L.}
\newblock \bibinfo{journal}{\bibinfo{title}{Uncovering the role of elementary
  processes in network evolution}}.
\newblock {\emph{\JournalTitle{Sci. Rep.}}} \textbf{\bibinfo{volume}{3}},
  \bibinfo{pages}{2920} (\bibinfo{year}{2013}).

\bibitem{conform-field-theor-3d}
\bibinfo{author}{Rychkov, S.}
\newblock \emph{\bibinfo{title}{{{EPFL Lectures}} on {{Conformal Field Theory}}
  in {{D}} {$\geq$} 3 {{Dimensions}}}}.
\newblock {{SpringerBriefs}} in {{Physics}} (\bibinfo{publisher}{{Springer}},
  \bibinfo{address}{{Geneva}}, \bibinfo{year}{2017}), \bibinfo{edition}{1st}
  edn.

\bibitem{facebook_gkbm10}
\bibinfo{author}{Gjoka, M.}, \bibinfo{author}{Kurant, M.},
  \bibinfo{author}{Butts, C.~T.} \& \bibinfo{author}{Markopoulou, A.}
\newblock \bibinfo{title}{Walking in facebook: A case study of unbiased
  sampling of osns}.
\newblock In \emph{\bibinfo{booktitle}{2010 Proceedings IEEE Infocom}},
  \bibinfo{pages}{1--9} (\bibinfo{organization}{Ieee}, \bibinfo{year}{2010}).

\bibitem{prefer-attach_kbm13}
\bibinfo{author}{Kunegis, J.}, \bibinfo{author}{Blattner, M.} \&
  \bibinfo{author}{Moser, C.}
\newblock \bibinfo{title}{Preferential attachment in online networks:
  Measurement and explanations}.
\newblock In \emph{\bibinfo{booktitle}{Proceedings of the 5th {{Annual ACM Web
  Science Conference}}}}, {{WebSci}} '13, \bibinfo{pages}{205--214}
  (\bibinfo{publisher}{{Association for Computing Machinery}},
  \bibinfo{address}{{New York, NY, USA}}, \bibinfo{year}{2013}).

\bibitem{bianconi-barabasi_bb01}
\bibinfo{author}{Bianconi, G.} \& \bibinfo{author}{Barab{\'a}si, A.-L.}
\newblock \bibinfo{journal}{\bibinfo{title}{Competition and multiscaling in
  evolving networks}}.
\newblock {\emph{\JournalTitle{EPL}}} \textbf{\bibinfo{volume}{54}},
  \bibinfo{pages}{436} (\bibinfo{year}{2001}).

\bibitem{quench-fitness_scb04}
\bibinfo{author}{Servedio, V. D.~P.}, \bibinfo{author}{Caldarelli, G.} \&
  \bibinfo{author}{Butt{\`a}, P.}
\newblock \bibinfo{journal}{\bibinfo{title}{Vertex intrinsic fitness: {{How}}
  to produce arbitrary scale-free networks}}.
\newblock {\emph{\JournalTitle{Phys. Rev. E}}} \textbf{\bibinfo{volume}{70}},
  \bibinfo{pages}{056126} (\bibinfo{year}{2004}).

\bibitem{taylor2015world}
\bibinfo{author}{Taylor, P.} \& \bibinfo{author}{Derudder, B.}
\newblock \emph{\bibinfo{title}{World city network: a global urban analysis}}
  (\bibinfo{publisher}{Routledge}, \bibinfo{year}{2015}).

\bibitem{grossman2017matching}
\bibinfo{author}{Grossman, G.~M.}, \bibinfo{author}{Helpman, E.} \&
  \bibinfo{author}{Kircher, P.}
\newblock \bibinfo{journal}{\bibinfo{title}{Matching, sorting, and the
  distributional effects of international trade}}.
\newblock {\emph{\JournalTitle{J. Polit. Econ.}}}
  \textbf{\bibinfo{volume}{125}}, \bibinfo{pages}{224--264}
  (\bibinfo{year}{2017}).

\bibitem{erdos-renyi_er59}
\bibinfo{author}{Erd{\H o}s, P.} \& \bibinfo{author}{R{\'e}nyi, A.}
\newblock \bibinfo{journal}{\bibinfo{title}{On random graphs}}.
\newblock {\emph{\JournalTitle{Publ. Math. Debr.}}}
  \textbf{\bibinfo{volume}{6}}, \bibinfo{pages}{290--297}
  (\bibinfo{year}{1959}).

\bibitem{ma2020role}
\bibinfo{author}{Ma, J.}, \bibinfo{author}{Valdez, L.~D.} \&
  \bibinfo{author}{Braunstein, L.~A.}
\newblock \bibinfo{journal}{\bibinfo{title}{Role of bridge nodes in epidemic
  spreading: {{Different}} regimes and crossovers}}.
\newblock {\emph{\JournalTitle{Phys. Rev. E}}} \textbf{\bibinfo{volume}{102}},
  \bibinfo{pages}{032308} (\bibinfo{year}{2020}).

\bibitem{crossover-sir-peak_mmb22}
\bibinfo{author}{Ma, J.}, \bibinfo{author}{Meng, X.} \&
  \bibinfo{author}{Braunstein, L.~A.}
\newblock \bibinfo{journal}{\bibinfo{title}{Peak fraction of infected in
  epidemic spreading for multi-community networks}}.
\newblock {\emph{\JournalTitle{J. Complex Netw.}}}
  \textbf{\bibinfo{volume}{10}}, \bibinfo{pages}{cnac021}
  (\bibinfo{year}{2022}).

\bibitem{power-law-heavy-tail_sp12}
\bibinfo{author}{Stumpf, M. P.~H.} \& \bibinfo{author}{Porter, M.~A.}
\newblock \bibinfo{journal}{\bibinfo{title}{Critical {{Truths About Power
  Laws}}}}.
\newblock {\emph{\JournalTitle{Science}}} \textbf{\bibinfo{volume}{335}},
  \bibinfo{pages}{665--666} (\bibinfo{year}{2012}).

\end{thebibliography}

\afterpage{\blankpage}
\afterpage{\blankpage}

\setcounter{page}{0}
\includepdf[pages=-]{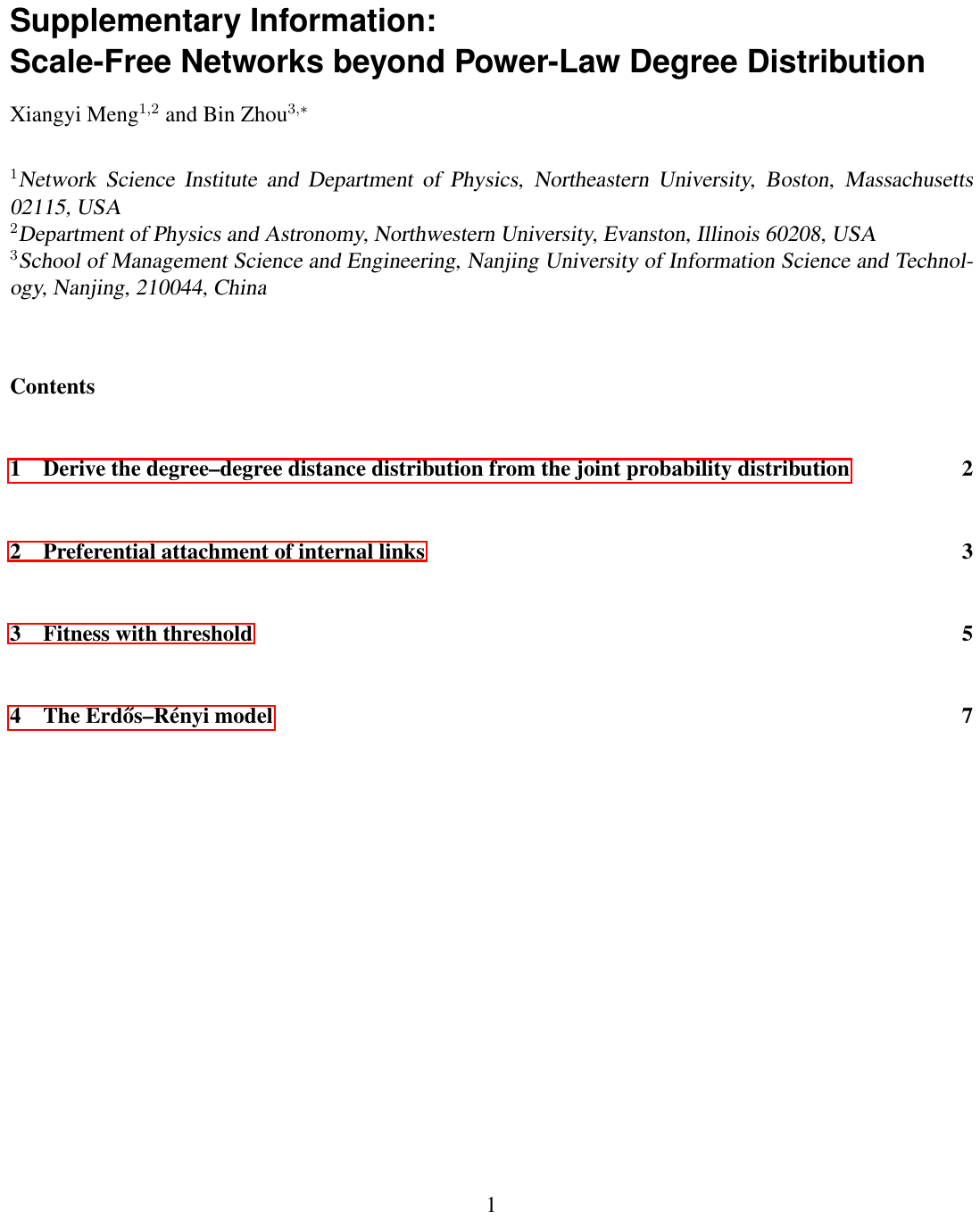}

\end{document}